\documentclass[12pt,fleqn]{article}  
\input{epsf}
\usepackage{amsmath}
\usepackage{latexsym}
\usepackage{amssymb}

\newlength{\dinwidth}
\newlength{\dinmargin}
\def\simle{\ \lower -2.5pt\hbox{$<$} \hskip-8pt \lower 2.5pt \hbox{$\sim$}\ }
\def\simge{\ \lower -2.5pt\hbox{$>$} \hskip-8pt \lower 2.5pt \hbox{$\sim$}\ }

\begin{document}
\newcommand{\diff}{{\rm d}}
\newcommand{\diffa}{{\rm d}^{3}}
\newcommand{\diffb}{{\rm d}^{4}}
\newcommand{\diffn}{{\rm d}^{n}}
\newcommand{\wt}{Ward-Takahashi identity}
\newcommand{\eu}{{\rm e}}
\newcommand{\im}{{\rm Im}}
\newcommand{\re}{{\rm Re}}
\newcommand{\tr}{{\rm tr}}
\newcommand{\sh}{\not\!}
\newcommand{\tp}{{\rm T}}
\newcommand{\vp}{{\rm P}}
\newcommand{\di}{{\rm Disc}}
\newcommand{\td}{triangular diagram}
\newcommand{\ab}{absorbitive part of the triangular diagram}
\newcommand{\dr}{dispersion relations}
\thispagestyle{empty}
\vspace*{-3cm}
\begin{flushright}
UPRF-00-10\\
June 2000
\end{flushright}
\vspace*{2.5 cm}

\begin{center}
{\huge Axial Anomaly Effects in Pion and $Z^0$ Radiative Decays}
\vskip 1.0cm
\begin{large}
{Luca~Trentadue and Michela~Verbeni}\\
\end{large}
\vskip .2cm
 {\it Dipartimento di Fisica, Universit\`a di Parma,\\
 INFN Gruppo Collegato di Parma, 43100 Parma, Italy}\\
\end{center}
\vspace*{.5cm}
 \begin{abstract}
We discuss a connection between axial anomaly and polarized radiative processes. By comparison with the corresponding unpolarized cases, we consider some physical outputs for the \(\pi^{+}\) and \(Z^0\) polarized radiative decays. We analyse in detail the pattern of mass singularity cancellation.
\end{abstract}
\vspace*{1cm}
\begin{center}
\end{center}

\newpage
\setcounter{page}{1}

\section{Introduction}\label{Intro}
Undoubtedly the axial anomaly represents a fundamental issue for understanding the basic aspects of quantum field theory. This issue has been analysed deeply over the years.\\
The anomaly problem has been treated by means of renormalization procedure, giving the interpretation of its origin in terms of ultraviolet divergences \cite{Adler}. A more formal analysis of the axial anomaly can be made by using the path integral formalism \cite{Fujikawa}.\\
Dolgov and Zakharov \cite{DZ} have shown an alternative approach to the axial anomaly, by studying the \(VVA\) triangular diagram through dispersion relations. From this approach follows the interpretation of the axial anomaly as an infrared phenomenon. It appears as due to a singularity present in the chiral limit in the absorbitive part of the triangular diagram.\\  
The infrared aspect of the axial anomaly, rised in this paper, is complementary to the more familiar ultraviolet one, which emerges from the renormalization procedure. A particularly interesting feature of this approach is that it allows to shed light upon the physical meaning of the anomalous chiral symmetry breaking, which is connected to a non conservation of helicity.\\
The connection between the anomaly and the breaking of a given symmetry has received a lot of attention in the literature and this subject has been discussed and developed in several papers. 
Gribov \cite{Gribov}, in a seminal work, has described the source of the anomalies as a collective motion of particles with arbitrarily large momenta in the vacuum. 
Related to this work, in ref. \cite{Mueller}, Mueller has discussed the manifestation of the axial anomaly as a flow of Landau levels.
In the papers of refs. \cite{Nielsen} the origin of the axial anomaly has been studied in two dimensions and, again, as a level crossing phenomenon.
The infrared interpretation of the axial anomaly, according to the Dolgov and Zakharov approach, has been possibly advanced in \cite{Falk}. In a series of papers \cite{S,Tung} the leading terms in the chiral limit have been correctly evaluated. Furthermore the dispersive analysis of the triangular \(VVA\) diagram is fundamental in the formulation of the 't Hooft consistency condition \cite{Hooft}.
The axial anomaly plays an essential role also in the interpretation of spin dependent parton distribution (see \cite{Carlitz}).\\
In this work we attempt to relate the Dolgov and Zakharov approach to the axial anomaly to some effects in the dynamics of physical reactions, as the radiative decays of \(\pi^{+}\) and \(Z^0\). We will try to show that the axial anomaly can be related to polarized radiative decays, as in the usual ultraviolet interpretation it is connected to the \(\pi^0 \to \gamma \gamma\) decay. We calculate the corresponding decay rates for the cases where the outgoing leptons are in a definite helicity state and we examine in some detail the structure of the mass singularities and their cancellation. We study how the Kinoshita-Lee-Nauenberg (KLN) theorem \cite{Lee} applies and we consider the analogies and the differences with respect to the corresponding unpolarized decay rates.\\
The paper is organized as follows. In the section \ref{Anomaly} we briefly reconsider the dispersive approach to the axial anomaly. In particular we concentrate on its physical origin.\\
In section \ref{helicity} we extend the Dolgov and Zakharov approach to the study of the radiative pion decay. We calculate the differential decay rate for the process, where the outgoing lepton undergoes an helicity flip and we interpret its behaviour in the chiral limit, as a manifestation of the axial anomaly.\\
In section \ref{Mass} we study the behaviour of the Inner Bremsstrahlung contribution to the pion decay rate in the collinear and infrared limits. We consider separately the unpolarized decay rate and both the cases of right-handed and of left-handed outgoing lepton. We find that the mass singularities cancellation mechanism occurs in different ways, according to the polarization of the outgoing lepton. We discuss the various realizations of the KLN theorem.\\
We also consider a more general process, i.e. the radiative \(Z^0\) decay in a lepton-antilepton pair, with a right-handed polarized lepton. We examine how the mass singularities cancel in this case and discuss the differences with respect to the pion decay.\\
Finally, in the Conclusions, we summarize our arguments.\\
A short discussion of the main results obtained has been already given in \cite{Trentadue}.

\section{The dispersive approach to axial anomaly}\label{Anomaly}

The Dolgov and Zakharov approach \cite{DZ} to derive the axial 
anomaly is based on the dispersion relation method. In this framework, the triangular diagram with two vector and one axial vertices is seen as the lowest order contribution to the process:
\begin{displaymath}
axial-vector~source \; \longrightarrow \; 2~real~photons,
\end{displaymath}
as described by the diagrams of fig. 1.
\begin{figure}
\begin{center}
\begin{tabular}{c}
\epsfxsize=19truecm
\epsffile{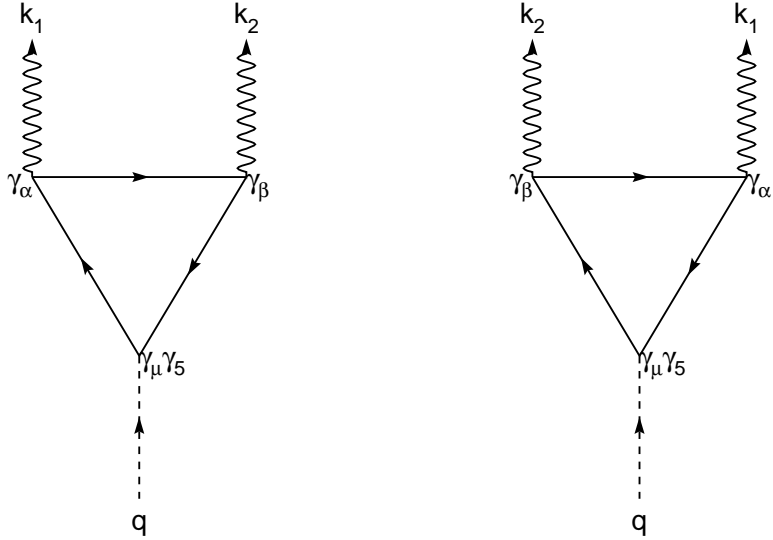}
\end{tabular}
\end{center}
\label{fig1}
\caption{Lowest order contribution to the process axial source $\to$ 2 photons.}
\end{figure}
In the physical region, the triangular diagram possesses a branch cut along the real axis, from \(4m^2\) to infinity. \(T^5_{\alpha\beta\mu}\) represents the corresponding amplitude.
We express it by means of a dispersion relation in the variable 
\(s=q^2=(k_1+k_2)^2\), where $k_1$ and $k_2$ are the outgoing photon momenta.\\
By requiring parity, Lorentz invariance, Bose symmetry and that \(T^5_{\alpha\beta\mu}\) satisfies the vector Ward-Takahashi identity,
\begin{displaymath}
k_1^{\alpha} T^5_{\alpha \beta \mu}(k_1,k_2)=
k_1^{\beta} T^5_{\alpha \beta \mu}(k_1,k_2)=0,
\end{displaymath}
we can write it in terms of an invariant scalar function \(g_1(q^2)\) as:
\begin{equation}
{\mit T}^{5}_{\alpha \beta \mu}(k_1,k_2) = \frac{2 \alpha}{\pi} g_1(q^2)\,
\epsilon_{\alpha \beta \sigma \rho}\, k_{1}^{\sigma} k_{2}^{\rho} q_{\mu}. 
\label{def1}
\end{equation} 
Similarly, we can express the contribution of the \td~with the vertex \(\gamma_{\mu} \gamma_5\) substituted by \(\gamma_5\) as:
\begin{equation}
{\mit T}^{5}_{\alpha \beta}(k_1,k_2) = \frac{\alpha m}{\pi} g_2(q^2)\, 
 \epsilon_{\alpha \beta \sigma \rho}\,k_{1}^{\sigma} k_{2}^{\rho}, \label{def2}
\end{equation}
where \(g_2(q^2)\) is another invariant scalar function.\\
\begin{figure}[t]
\begin{center}
\begin{tabular}{c}
\epsfxsize=11truecm
\epsffile{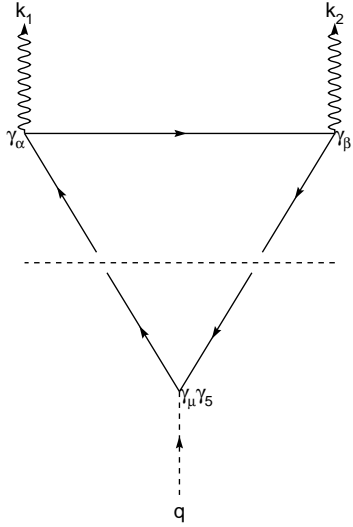}
\end{tabular}
\end{center}
\label{fig2}
\caption{Cut diagram}
\end{figure}
In terms of dispersion relations, the functions \(g_i(q^2)\), \(i=1,2\) can be expressed as follows:
\begin{equation}
g_i(q^2) = \frac{1}{\pi} \int^{+\infty}_{4m^2} \diff s \, 
\frac{\im \, g_i(s)}{s-q^2}\;\;\;\;\;\;\;i=1,2.
\label{dr}
\end{equation}
We can use unsubtracted dispersion relations, because the integrals contained in \(g_i(q^2)\), \(i=1,2\) are convergent, since these functions are multiplied by three 
and two powers of momentum, respectively.\\
The imaginary part of the invariant scalar functions can be derived from the absorbitive part of the \td, calculated by cutting the diagram as shown in fig. \ref{fig2}
and by using the Cutkosky rules or the perturbative unitarity relation. We obtain \cite{DZ}:
\begin{eqnarray}
\im g_1(q^2) & = & - \frac{2 \pi m^2}{q^4} \theta(q^2 - 4 m^2)
\ln \left( \frac{1 + \sqrt{1 - 4 m^2/q^2}}{1 - \sqrt{1 - 4 m^2/q^2}}\right).
\label{g1}\\
\im g_2(q^2) & = & - 2 \pi \theta(q^2 - 4 m^2) \frac{1}{q^2}
\ln \left( \frac{1 + \sqrt{1 - 4 m^2/q^2}}{1 - \sqrt{1 - 4 m^2/q^2}} \right)
 \label{g2}
\end{eqnarray}  
Using eqs. (\ref{def1}), (\ref{def2}), (\ref{dr}) and the above expressions, we derive the complete \td~contribution:
\begin{equation}
{\mit T}^{5}_{\alpha \beta \mu}(k_1,k_2) = \frac{2 m}{q^2 + i \epsilon} \,
{\mit T}^{5}_{\alpha \beta}(k_1,k_2) q_{\mu} +
\frac{2 \alpha }{\pi} \frac{1}{q^2 + i \epsilon}\, \epsilon_{\alpha \beta \sigma \rho} \,
                 k_{1}^{\sigma}k_{2}^{\rho} q_{\mu} \label{tr}
\end{equation}
and the anomalous axial-vector \wt:
\begin{equation}
q^{\mu} {\mit T}^{5}_{\alpha \beta \mu}(k_1,k_2) = 
\frac{2 m}{q^2}\, {\mit T}^{5}_{\alpha \beta}(k_1,k_2) +
\frac{2 \alpha}{\pi}\, \epsilon_{\alpha \beta \sigma \rho} \, k_{1}^{\sigma}k_{2}^{\rho} 
.
\label{awt}
\end{equation}
Thus, in the dispersive approach we find the same result obtained by 
using the renormalization procedure \cite{Adler}.
The method above allows a more direct treatment, since we avoid
evaluating divergent integrals and introducing regularization schemes.\\
The fact that the axial anomaly can be derived without using the renormalization procedure, suggests that this should not be considered as the only origin of the anomalous breaking of the chiral symmetry.
Moreover, by studying the axial anomaly with the renormalization procedure, that is by considering its ultraviolet interpretation, an important aspect of this phenomenon remains obscure and we are bound by a formal derivation only.
As stressed in \cite{DZ} and \cite{Z}, the dispersive approach shows that the anomaly is related to the chiral limit and therefore, it can be interpreted as an infrared effect. In this work we are close to this infrared interpretation, which, as we will see, can help to shed light on some aspects of the physics connected to the axial anomaly.\\
Let us briefly discuss the physics involved in the amplitudes contributing to the absorbitive part of the triangular diagram. The two Born diagrams, obtained from the cut of the triangular diagram (see fig. \ref{fig1}), describe the following two processes:
\begin{enumerate}
\item[a)] the production of a fermion-antifermion pair (for example \(e^{+}e^{-}\)) by an axial-vector source;
\item[b)] the subsequent annihilation of the pair into two real photons.
\end{enumerate} 
In both these processes there occur helicity flips, thus the chirality is not conserved in the zero mass limit.\\
Let us go to the center of mass frame of the two final photons.
 In the first process the axial-vector source produces an \(e^{+}e^{-}\) pair of total spin zero, since a spin 1 
state cannot annihilate into a two real photons state. Thus \(e^{+}\) and \(e^{-}\) must have the same helicity and hence opposite chirality in the massless limit\footnote{ Equivalently, considering the outgoing antifermion as an incoming fermion, one can say that the latter makes an helicity flip interacting with the axial-vector source.}.
In the process b) the \(e^{+}e^{-}\) pair annihilates into two real photons by going through an intermediate virtual state.
There are four possible virtual states \cite{Huang}: 
one is drawn in fig. 3 
\begin{figure}[t]
\begin{center}
\begin{tabular}{c}
\epsffile{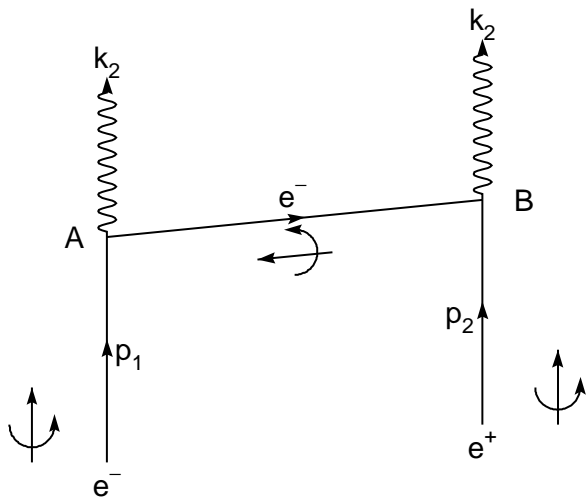}
\end{tabular}
\end{center}
\label{fig3}
\caption{Process b)}
\end{figure}
(\({\bf p}_1\) and \({\bf p}_2\) are the \(e^{-}\) and \(e^{+}\) linear momenta), a second one is obtained by reversing the virtual state helicity; the remaining two are obtained replacing the outgoing virtual fermion with an incoming virtual antifermion. Let us study the case shown in fig. 3 and assume that \(e^{+}\) and \(e^{-}\) are both right-handed. In the vertex B the chirality is conserved in the massless limit, while in the vertex A there is an helicity flip, thus the chiral symmetry is broken. As can be easily checked, for all the remaining virtual states we always have an allowed vertex and a forbidden one. \\
At the Born level, these reactions are described by the classical QED Lagrangian, which, in the \(m \to 0\) limit, is invariant under chiral transformations. 
Thus it seems at first sight that the \ab, being proportional to the product of amplitudes relative to processes forbidden by chiral invariance, vanishes. On the contrary, one sees that taking the limit \(m \to 0\) in (\ref{g2}) gives a finite result \cite{DZ}:
\begin{equation}
\im g_1(q^2)\,\longrightarrow \, - \pi \delta(q^2)\;\;\;\;\;\;\;{\rm as}~m\rightarrow0.
\label{limit}
\end{equation}  
Therefore, by studying the \ab, one establishes that a non-conservation of helicity occurs, becoming, in the massless limit, a non-conservation of chirality. We interpret this as related to the presence of the anomalous term in the divergence of the axial-vector current. Thus the axial anomaly can be derived by studying the properties of the amplitude in the infrared region.\\
Even if in this work we will analyse the cancellation of mass singularities in polarized processes, we will not discuss the physical implication of the zero fermion mass limit.\\
As stated in refs. \cite{DZ,S,Z}, the result in eq. (\ref{limit}) indicates that there occurs a cancellation of the suppression factor \(m^2\), due to the terms coming from the vertices with helicity flip. In (\ref{g1}) the 
logarithmic factor conspires to give a finite result. This logarithm is a collinear one; we shall discuss about this kind of logarithms in section \ref{Mass}. Its presence is a manifestation of the singularity occurring in the fermion propagator as \(m \to 0\), which exactly cancels the suppression factor \(m^2\) in the numerator.\\
We observe that the behaviour of the \ab~given in eq. (\ref{limit}) shows that the \(m \to 0\) limit is not smooth; if we evaluate the amplitudes with the massless theory, they identically vanish. If, on the contrary, we take the chiral limit after summing their product over the intermediate states, we obtain a result different from zero.\\

\section{Helicity changing processes and polarized pion decay}\label{helicity}
\subsection{Helicity changing processes}
In the process b) contributing to the \ab~a charged polarized fermion, changes helicity by emitting a photon. Thanks to the collinear singularity within the propagator relative to this fermion, the chiral suppression factor $m^2$ is cancelled and the \ab~ doesn't vanish in the chiral limit.\\
Let us now extend these remarks to physical processes having the same features as the ones characterizing the \ab. This means that we want to investigate about a manifestation of the axial anomaly in reactions where a fermion changes helicity by emitting a photon. We call these processes helicity changing processes. They have been considered by Lee and Nauenberg \cite{Lee}, who observed that states with opposite helicity don't decouple in the massless limit, provided that this limit is taken after having summed the transition probability over the final phase space. In the process b) the incoming \(e^{+}\) and \(e^{-}\) are in a definite helicity state, as discussed above. In the reactions we want to study, we calculate the probability that an outgoing fermion assumes an helicity opposite to the one required by the interaction before the emission of the photon. In other words, we evaluate the probability that in the fermion-photon vertex occurs an helicity flip. 
According to the Dolgov and Zakharov analysis, we interpret the presence of a term independent of the fermion mass in the corresponding cross sections as a manifestation of the axial anomaly. Due to this term, the probability for a process with helicity flip doesn't vanish in the chiral limit.\\

\subsection{Polarized radiative pion decay}\label{pionpol}
We first examine the non radiative pion decay
\begin{displaymath}
\pi^{+} \, \longrightarrow\, l^{+} + \nu_l, 
\end{displaymath}
where \({l^{+}}\) is an antilepton (\({e^{+}}\) or \({\mu^{+}}\)) and \({\nu_l}\) is the associated neutrino. At the Born level the total decay rate is given by:
\begin{equation}
\Gamma_0(\pi^{+}\rightarrow l^{+} \nu_{l}) = 
\frac{G^2 f^{2}_{\pi}}{8 \pi}\, \mid V_{ud}\mid^2 \, \frac{m^2_l}{m^3_{\pi}}\,
(m^2_{\pi} - m^2_l)^2,\label{born}
\end{equation}
where \(G\) is the Fermi coupling constant, \(f_{\pi}\) is the pion decay constant, \(V_{ud}\) is the CKM matrix element, \(m_l\) and \(m_{\pi}\) are the lepton and pion masses, respectively.
The decay rate (\ref{born}) is proportional to $m^2_l$, since, due to angular momentum conservation, the pion produces a left-handed lepton, while the structure of the weak coupling requires the $l^+$ to be right-handed for \(m_l=0\).\\
This situation is confirmed by the expression of the total decay rate for the
process in which the antilepton is polarized:
\begin{eqnarray}
\Gamma^{pol}_{0}(\pi^{+}\rightarrow l^{+} \nu_l) = 
\frac{G^2 f^{2}_{\pi}}{16 \pi} \, \mid V_{ud} \mid^2 \, \frac{m^2_l}{m^3_\pi}\,
                     (m^{2}_{\pi} - m^{2}_{l})^2 \,
                     \left(1 - \frac{{\bf p}_l \cdot {\bf s_{l}}}
                    {|{\bf p}_{l}|} \right), 
                       \label{bornpol}
\end{eqnarray}
where \({\bf p}_{l}\) and \(\bf s_l\) are the linear momentum and the spin vector of the lepton, respectively, giving \({\bf p}_l \cdot {\bf s}_l/|{\bf p}_{l}| = \pm 1 \), for right-handed and left-handed lepton, respectively.
The eq. (\ref{bornpol}) indicates 
that the lepton is mandatory left-handed, as requested by angular momentum conservation.\\ 
We now consider the radiative correction, due to the emission of a real photon, to the pion decay,  
\begin{displaymath}
\pi^{+} \rightarrow l^{+} + \nu_{l} + \gamma,
\end{displaymath}
when the outgoing antilepton is polarized, i.e. when it is in a definite 
helicity state. As we have seen in the non-radiative case, due to angular momentum conservation, the \(\pi^{+}\) is coupled to a left-handed lepton. We calculate the probability that the lepton flips its helicity and becomes right-handed, by emitting a real photon. \\
The amplitude describing the radiative pion decay can be divided into two parts, the Inner Bremsstrahlung and the Structure Dependent amplitudes \cite{Bryman}:
\begin{equation}
M(\pi^{+}\rightarrow l^{+} \nu_l \gamma) = M_{IB} + M_{SD}. \label{e:4.4}
\end{equation}
The Inner Bremsstrahlung amplitude, where the photon is radiated 
from the external charged particles, can be calculated using the rules of QED, with a point like pion coupling; the Structure-Dependent 
amplitude is governed by the strong interactions. \\
Clearly, the relevant part for the problem we are considering is the Inner 
Bremsstrahlung contribution described by the diagrams of fig. 4.
\begin{figure}[t]
\begin{center}
\begin{tabular}{c}
\epsffile{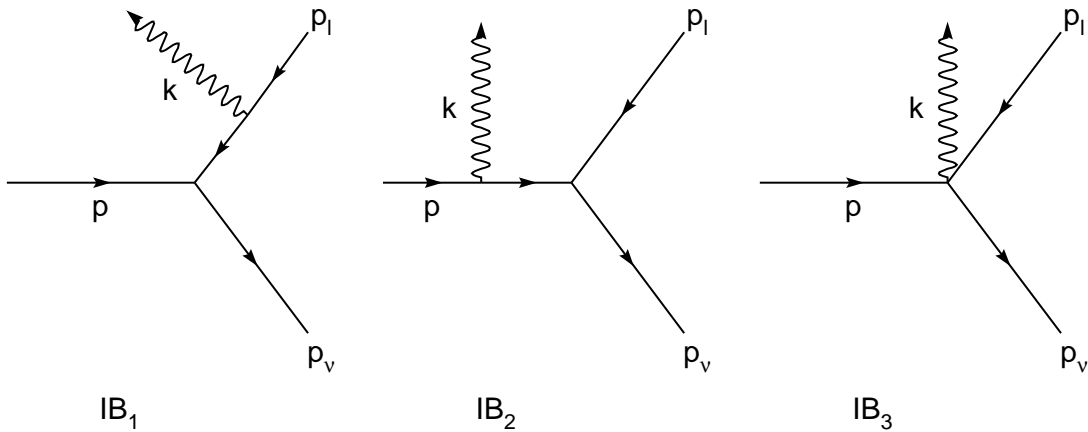}
\end{tabular}
\end{center}
\label{innerb}
\caption{Inner Bremsstrahlung diagrams}
\end{figure}                   
The term associated to the \(IB_3\) diagram is the so called contact term and 
it is introduced to ensure gauge invariance (see, for example, \cite{Bryman}).\\
We consider only tree level diagrams, since, to reveal the effects of the axial anomaly, it is sufficient to take into account the contribution of the vertex with the helicity flip. For the moment, we may neglect the corrections due to the emission of 
virtual photons; these will be discussed in section \ref{Mass}.
Finally, at this order in perturbation theory, we retain terms of all powers 
in the lepton mass. We will argue that the manifestation of the axial
anomaly is strictly connected to these terms.\\
The Inner Bremsstrahlung amplitude is given by \cite{Bryman}:
\begin{eqnarray}
M_{IB} & = & \sum_{i=1}^{3}\, IB_{i}  \nonumber \\
       & = & i e \, \frac{G}{\sqrt{2}}\, f_{\pi} \, V_{ud}\, m_l\,
              \bar{u}(p_{\nu}) (1 + \gamma_5) 
          \left[\frac{p \cdot \epsilon}{p \cdot k} - 
                \frac{\not\!k \not\!\epsilon + 2 p_l \cdot \epsilon}
                     {2 p_l \cdot k} \right] v(p_l, s_{lR}), \label{e:4.5}
\end{eqnarray}
where \(u\) and \(v\) are the Dirac spinors for the neutrino and the lepton, respectively, \(p\) is the pion momentum, \(p_l\) and \(s_{lR}\) are the momentum and the polarization vector of the lepton, \(k\) and 
\(\epsilon\) are the momentum and the polarization vector of the 
photon.\\
 We see that \(M_{IB}\) is proportional to the lepton mass \(m_l\), thus the 
decay rate is proportional to \(m^2_l\). As we have said above, this factor is a consequence of the structure of the weak coupling. Thus, if
we remove it by normalizing the radiative decay rate with respect to the 
non radiative one, we can emphasize the mass dependence of the radiation 
emission process.\\
The differential Inner Bremsstrahlung 
contribution for right-handed lepton is:
\begin{eqnarray}
& &\frac{1}{\Gamma_0}\,\frac{\diff \Gamma^{R}_{IB}}{\diff y} =
\frac{\alpha}{4 \pi}\,\frac{1}{(1-r)^2} \, \frac{1}{A(1-y+r)}
\bigg\{2A \, \Big[1+y^2 - 2A + r(2A+r-6)\Big] \nonumber \\
& & + \,\Big[(A+2r)(1+r)^2 + Ay(y-4r) + 2ry(1+y)-y(1+y^2+5r^2)\Big]\,\ln{\frac{y+A}{y-A}} \nonumber \\
& & + \,(1-y+r)^2(y-2r-A) \ln{\frac{y+A-2}{y-A-2}}\bigg\}.
\label{IBR}
\end{eqnarray}
The dimensionless variable \(y\) is defined as 
\begin{displaymath}
y \, =\, \frac{2E_l}{m_{\pi}}
\end{displaymath}
and 
\begin{displaymath}
r \, =\, \frac{m^2_l}{m^2_{\pi}},\;\;\;\;\;\;A \, =\,\sqrt{y^2-4r}.
\end{displaymath}
The physical region for \(y\) is:
\begin{equation}
2\sqrt{r} \le y \le 1+r.
\label{region}
\end{equation}
We see that, in eq. (\ref{IBR}), there is a term independent of the lepton mass, the one in the first square brackets. Owing to this term, the differential decay rate 
doesn't vanish in the limit \(m_l \rightarrow 0\). We have:
\begin{equation}
\frac{1}{\Gamma_0} \frac{\diff \Gamma^R_{IB}}{\diff y}\,
\longrightarrow \, \frac{\alpha}{2 \pi} \, (1-y)\;\;\;\;\;\;\;\;{\rm as}\;r \to 0.\label{massa0}
\end{equation}
This indicates that there occurs an helicity flip and thus a chirality non conservation in the limit of zero lepton mass. According to the interpretation given above, 
this term can be interpreted as connected to the axial anomaly. It corresponds to the anomalous term present in the divergence of the axial current.\\
Since the polarized radiative process with the right-handed $l^+$ is forbidden, in the limit
 \(m_l \rightarrow 0\), by the chiral invariance of the massless QED
Lagrangian, the appearance of a term different from zero, in this limit, indicates the action of a cancellation mechanism, analogous to the one acting in the absorbitive part of the triangular diagram.\\ 
To see how this mechanism acts, let us examine  
the decay rate differential with respect to the lepton energy and to the emission angle:
\begin{eqnarray}
& & \frac{1}{\Gamma_0}\, \frac{\diff \Gamma^R_{IB}}{\diff y \,\diff \cos{\theta}}
\, = \, \frac{\alpha}{4 \pi}\,\frac{1}{(1-r)^2} \, \frac{1}{(1-y+r)}
\bigg\{\frac{4r}{(y-A\cos{\theta})^2} \, (1+y^2-y-A) \nonumber \\
& & + \frac{4r^2}{(y-A\cos{\theta})^2}(A+r-y-2) + y(1+r) - A(1-r) - 4r \nonumber \\
& & + \, \Big[(A+2r)(1+r)^2 + Ay(y-4r) + 2ry(1+y) -y(1+y^2+5r^2)\Big]
\, \frac{1}{(y-A\cos{\theta})} \nonumber \\
& & + (1-y+r)^2(y-A-2r)\, \frac{1}{(y - A\cos{\theta} -2)} \bigg\}.
\label{integrale}
\end{eqnarray} 
The first term is proportional to the lepton propagator squared. Thanks to this term, when we
carry out the final integration over the emission angle, we obtain, besides 
logarithmic collinear divergences, also power collinear divergences. These power terms are essential for the convergent behaviour of the distribution.
The cancellation of the chiral suppression would not take place were these terms absent. The term related to the axial anomaly (see eq. (\ref{massa0})) originates from this cancellation. To obtain the differential decay rate, we have to evaluate integrals of the form:
\begin{eqnarray}
m^2_l \int_{-1}^{+1} \diff \cos{\theta} \: \frac{1}
        {(E - \sqrt{E^2 - m^2_l} \cos{\theta})^2} = \nonumber \\
         \frac{m^2_l}{\sqrt{E^2 - m^2_l}}\, \frac{2 \sqrt{E^2 - m^2_l}}{m^2_l}, 
\label{e:4.7}
\end{eqnarray} 
where the propagator has a power mass singularity that exactly cancels the 
factor \(m^2_l\), coming from the vertex lepton-photon.\\
 The integration of the terms in equation (\ref{integrale}) containing the lepton propagator gives rise to the collinear logarithms:  
\begin{equation}
\ln{\frac{E + \sqrt{E^2 - m^2_l}}{E - \sqrt{E^2 - m^2_l}}} \simeq
\ln{\frac{m_l}{E}}. \label{e:4.8}
\end{equation}
In the differential decay rate (\ref{IBR}) there are also terms proportional to the logarithm
\begin{equation}
\ln{\frac{E + \sqrt{E^2 - m^2_l} - m_{\pi}}
         {E - \sqrt{E^2 - m^2_l} - m_{\pi}}}. \label{e:4.9}
\end{equation}
This one is another collinear logarithm; it diverges 
in the limit \(m_{\pi} \rightarrow 0\) and \(m_l \rightarrow 0\) and 
corresponds to the possibility that the photon is emitted parallel to the
 pion.\\
Finally, we point out that the chiral limit (\(m_l \rightarrow 0\)) is not smooth. In fact we get different results depending upon whether we 
describe an helicity changing process using the massless theory or we take
the \(m_l \rightarrow 0\) limit, after carrying out the integration over the final phase space. The radiative pion decay is not a good process to see this, since, owing to the angular momentum conservation in the pion vertex, this process cannot take place within a massless theory, given the structure of the \((V-A)\) coupling of the electroweak theory.
To see that the chiral limit is not a smooth one, we consider the scattering of a polarized electron with a proton (treated as a point like particle) of initial and final momenta \(q\) and \(q'\) respectively, accompanied by the emission of a real photon. We consider a left-handed incoming electron with momentum \(p\) and spin \(s_L\): we calculate the probability that the electron makes an helicity flip, emitting a real photon with momentum \(k\) and polarization $\epsilon$ and thus becoming right-handed electron with momentum \(p'\) and spin \(s_R\):
\begin{displaymath}
e(p,s_L) + p(q) \to e(p',s_R) + p(q') + \gamma(k).
\end{displaymath}
We study this process with the massless QED. The left-handed and right-handed spinors are given respectively by:
\begin{eqnarray*}
u(p,s_L) & = & \frac{1 - \gamma_5}{2}u(p)  \\
u(p',s_R) & = & \frac{1 + \gamma_5}{2} u(p').
\end{eqnarray*}
The corresponding transition amplitude identically vanishes:
\begin{eqnarray}
M(e_lp \to e_Rp \gamma) & = & \frac{e^3}{[(p'+k)^2 + i \epsilon]
(l^2 + i \epsilon)} \nonumber \\
& & \bar{u}(p')\frac{1-\gamma_5}{2} \sh\epsilon(\sh p' + \sh k)
\gamma_{\rho} \frac{1-\gamma_5}{2}u(p)
\bar{u}(q')\gamma^{\rho}u(q) = 0,
\end{eqnarray} \label{QED0}
since it contains the product of different chirality projectors.\\
We now calculate the cross sections for processes with helicity flip using the massive QED and then we take the massless limit after having summed the transition probability over the final phase space. An example can be found in \cite{Falk}, where the cross section for the process
\begin{equation}
e^{-}(p_{-},\lambda) + A(p)\,\longrightarrow\, e^{-}(p'_{-
},\lambda') +
\gamma(k,\lambda_{\gamma}) + B(q_i) \label{procr}
\end{equation}
is calculated. Here \(A\) is the target (for example another fermion), \(\gamma\) is a bremsstrahlung photon, assumed almost collinear with respect to the direction of the incident electron and \(B\) is a set of particles produced in the reaction. \\
The helicity flip cross section is given by: 
\begin{equation}
\frac{\diff \sigma_{hf}}{\diff x} = \sigma_0 \left(s(1-x) 
\right)
                     \frac{\alpha}{2 \pi} x, 
 \label{hf}
\end{equation}
where \(x = k_0/E\), \(E\) is the energy of the incoming electron and \(\sigma_0\) is the cross section for the Born process
\begin{equation}
e^{-}(p_{-} - k,\lambda) + A(p)\,\longrightarrow\,e^{-}(p'_{-
},\lambda') +
B(q_i). \label{procnr}
\end{equation}
We see that the expression (\ref{hf}) does not vanish in the massless limit.\\
The result in eq. (\ref{hf}) coincides with the one in eq. (\ref{massa0}) at leading order. Indeed, in \cite{Falk} the subdominant terms are not accounted for. As it will become apparent in the following section, these terms are essential for the cancellation of mass singularities.\\
In ref. \cite{Einhorn}, the authors give a different interpretation of the helicity changing processes and hence come to a different conclusion about the smoothness of the zero mass limit.\\

\section{Mass singularities}\label{Mass}
As it is well know, there are two types of divergences occurring in a theory
 when the mass of a particle goes to zero, which will be comprehensively call in the following mass singularities.\\
 The first type of divergences appears when we reach the phase space region, where the momentum of the massless 
particles vanishes: these are called infrared divergences. They occur for example in QED when the energy of the photon goes to zero. The Block-Nordsieck theorem \cite{Bloch} assures that the infrared divergences cancel out in any inclusive cross section. \\
The other type of mass singularities occurs in theories with massless coupled particles, like in QED when the photon couples to a fermion, in the limit of zero fermion mass. The origin is purely kinematical: when two massless particles, say with momenta \(k\) and \(k'\), move parallel to 
each other, they have combined invariant mass equal to zero:
\begin{equation}
q^2 = (k+k')^2 = 2EE'(1-\cos\theta) \rightarrow 0\;\;\;\;\;\; {\rm as}\; \theta \rightarrow 0,
\end{equation}
even though neither \(k\) nor \(k'\) are soft. These divergences are called collinear singularities.\\
If we keep the fermion mass finite and integrate 
over the photon emission angle, the collinear divergence doesn't occur, but the possibility of a divergence in the limit \(m \rightarrow 0\) results in the presence of the collinear logarithms, that is logarithms of the form 
\(\ln{(E/m)}\), diverging for \(m \to 0\).\\
In the case of collinear singularities, the theorem by Kinoshita, Lee and 
Nauenberg \cite{Lee} guarantees that these divergences cancel out if we sum the transition probability over the set 
of degenerate states, order by order in perturbation theory. This cancellation mechanism is analogous to that of the infrared divergences, 
as stated by the Block-Nordsieck theorem. Both types of mass singularities arise because the states of a theory with massless particles are highly degenerate. The infrared divergences 
can be interpreted as a consequences of the fact that a state with a single charged particle is degenerate with a state made of the same particle plus a number of soft photons; this correspond to the impossibility of distinguishing experimentally a charged particle from one accompanied by soft photons, owing to the finite resolution of the measurement apparatus. The situation of the 
collinear singularities is analogous: the state with a massless charged particle is degenerate 
with the states containing the same particle and a number of collinear photons. This corresponds to the fact that, as a consequence of the finite angular resolution, we cannot establish if a massless charged particle is accompanied by collinear photons. \\
Let us now discuss the structure of mass singularities and their cancellation in the decay rates for the radiative pion and $Z^0$ decays and how the KLN theorem applies to this cases.\\

\subsection{The pion case}\label{pion}
It is useful to separate the cases of unpolarized, right-handed and left-handed outgoing lepton. \\
Let us consider first the mass singularity cancellation mechanism for the familiar case of unpolarized radiative 
\(\pi^{+}\) decay to the first order in $\alpha$. The differential Inner Bremsstrahlung contribution is given by:
\begin{eqnarray}
\! \frac{1}{\Gamma_0}\,\frac{\diff \Gamma_{IB}}{\diff y} & = & 
\frac{\alpha}{4 \pi}\,\frac{1}{(1-r)^2} \, \frac{1}{(1-y+r)}\,
\bigg\{4A(r-1) + \Big[(1+r)^2 + y(y-4r)\Big]\,\ln{\frac{y+A}{y-A}} \nonumber \\
& - & (1-y+r)^2\,\ln{\frac{y+A-2}{y-A-2}} \bigg\}.
\label{IRnp}
\end{eqnarray}
One can easily see that the eq. (\ref{IRnp}) is divergent both in the collinear and in the infrared limits. The coefficients of the collinear logarithms don't go to zero in the limit \(r \rightarrow 0\). There are also infrared divergences, because if we let \(y\) reach its kinematical limit \(y^{MAX} = 1+r\),
corresponding to the photon energy going to zero, the expression (\ref{IRnp}) diverges.\\
The decay rate is made free from mass singularities in the ordinary way: the divergences cancellation occurs 
in the total inclusive decay rate, when we add all the first order contributions to the 
perturbative expansion, i.e. those relative to real and virtual photon emission.\\
The diagrams describing the real photon emission contribution were already given in fig. 4; the 
diagram for the virtual correction is drawn in fig. 5.
\begin{figure}[t]
\begin{center}
\begin{tabular}{c}
\epsffile{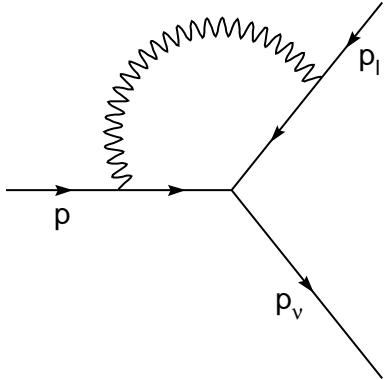}
\end{tabular}
\end{center}
\label{fig5}
\caption{Virtual diagram}
\end{figure}
The expression for the lepton energy spectrum, including the Inner Bremsstrahlung contribution and the virtual photon one, calculated to the leading order in \(m_l/m_{\pi}\), is given by \cite{Kuraev}:
\begin{equation}
\frac{1}{\Gamma_0} \frac{\diff \Gamma}{\diff y} = 
D(y,r)[1 + \frac{\alpha}{\pi} K_l(y)].  \label{spettroy}
\end{equation} 
\(D(y,r)\) is the lepton distribution function, given, to the first order in \(\alpha\), by:
\begin{equation}
D(y,r) = \delta (1-y) + \left[ \frac{\alpha}{\pi}(L-1) + O(\alpha^2)\right]P^{(1)}(y),
\label{defD}
\end{equation}
where \(L\) is the logarithm
\begin{equation}
L = \ln{\frac{m_{\pi}}{m_l}},
\label{defL}
\end{equation}
diverging in the collinear limit; if the lepton is an electron, \(L \simeq 5.6\).
\(P^{(1)}(y)\) is the Gribov-Lipatov-Altarelli-Parisi kernel \cite{Altarelli}, which can be expressed in the form:
\begin{equation}
 P^{(1)}(y) = \frac{1+y^2}{1-y} - \delta (1-y) \int^1_0 \diff z 
\frac{1+z^2}{1-z}.
\label{defP}
\end{equation}
\(K_l(y)\) is a finite term, free from infrared and collinear singularities, which has the expression:
\begin{equation}
K_e(y) = 1 - y - \frac{1}{2}(1-y)\ln{(1-y)} +
\frac{1+y^2}{1-y}\ln{y}.
\label{defK}
\end{equation} 
 The differential decay rate to order \(\alpha\) therefore becomes:
\begin{equation}
\frac{1}{\Gamma_0}\, \frac{\diff \Gamma}{\diff y} = 
\delta(1-y) + \frac{\alpha}{\pi}\,(L-1)\,P^{(1)}(y) + 
\frac{\alpha}{\pi}\,K_e(y).
\label{ordalfa}
\end{equation}
To calculate the inclusive decay rate, we have to integrate the expression in eq. (\ref{ordalfa}) over \(y\); to the leading order in the lepton mass, the physical region for \(y\) is \(0 \le y \le 1\). The kernel \(P^{(n)}\) has the property that:
\begin{equation}
\int^1_0 \diff y P^{(n)}(y) = 0;
\label{canc}
\end{equation}
thus, when we calculate the inclusive decay rate, the coefficient of the collinear logarithm vanishes and the resulting expression is finite in the zero mass limit.\\
Carrying out the integration over \(y\), we obtain the well known inclusive decay rate to order \(\alpha\)
\begin{equation}
\frac{\Gamma}{\Gamma_0} = 1 + 
\frac{\alpha}{\pi}\,[\frac{15}{8} - \frac{\pi^2}{3}].
\label{tot}
\end{equation}
As expected, the expression (\ref{tot}) is finite in the collinear limit and is also free from infrared divergences, because, as usual, the infrared divergences present in the soft photon contribution and in the virtual photon contribution have cancelled each other.\\
Let us now discuss the mass singularities in the case of the right-handed Inner Bremsstrahlung contribution, given in eq. ({\ref{IBR}}). It is easy to see 
that \(\diff \Gamma^R_{IB}/\diff y\) is finite in the limit \(r \to 0\), 
i.e. it is free from collinear singularities. In this limit the coefficients 
of both the collinear logarithms vanish. Indeed, as we have observed in section \ref{pionpol}, only the term related to the axial anomaly survives in the zero mass limit, that is the part of the decay rate independent of 
the lepton mass.\\
We observe that the right-handed Inner Bremsstrahlung contribution is free from infrared divergences as well. If we make the lepton energy 
\(y\) reach its kinematical limit \(y^{MAX}\), we obtain a finite result:
\begin{eqnarray}
\frac{1}{\Gamma_0}\frac{\diff \Gamma^R_{IB}}{\diff y} \longrightarrow  0\;\;\;\;\;\;\;{\rm as}\;y \to y^{MAX}.
\label{limIR}
\end{eqnarray}
The result ({\ref{limIR}}) shows that the soft photon emission does not contribute to the radiative \(\pi^{+}\) decay with a right-handed lepton. This is a consequence of the fact that the soft photon contribution factorizes with respect to the Born decay rate, but this vanishes in the case of right-handed $l^+$ (see eq. (\ref{bornpol})). Physically, eq. (\ref{limIR}) is due to the fact that soft photons don't carry spin, thus they cannot contribute to the angular momentum balance; therefore the process with the right-handed lepton emitting a soft photon is forbidden by angular momentum conservation.\\
For the same reason of angular momentum conservation, in the right-handed case also the virtual contribution identically vanishes. The virtual photon diagram (see fig. 5) interferes with the Born one; the corresponding correction to the total decay rate for \(\pi \to l \nu_l\) was calculated long ago by Kinoshita \cite{Kinoshita59} and is given by:
\begin{eqnarray}
\Gamma_v = \frac{\alpha}{2\pi}\,\bigg\{3 \ln{\frac{\Lambda}{m_{\pi}}} - \frac{1}{2} b(r)\,\Big[4\ln{\frac{\lambda}{m_{\pi}}} - \ln{r} + 3
\Big]  + \frac{r}{1-r} \ln{r} + 1\bigg\}\Gamma_0,
\label{kin}
\end{eqnarray}
where
\begin{displaymath}
b(r) = \frac{1+r}{1-r}\,\ln{r} + 2.
\end{displaymath}
$\Lambda$ is the ultraviolet cutoff and $\lambda$ is the infrared one.\\
As usual, the virtual correction is factorized with respect to the Born decay rate, but, as we have already seen, if the lepton is right-handed, this is identically 
zero.\\
In the right-handed case, the mass singularities cancellation occurs trough a mechanism different from the one working in the unpolarized decay rate. The infrared and the collinear limits give separately a finite result. In particular, the coefficient of the collinear logarithms is the lepton mass, instead of the usual correction factor coming from the soft and the virtual photon contributions, as in eq. (\ref{ordalfa}). In this sense, since the soft and collinear radiation factorizes with respect to the Born helicity changing decay rate, the double logarithm Sudakov term can be equally factorized. It could be useful to investigate the impact of the higher order terms on the radiative correction to the Born amplitude. It is an open question if say hard collinear photons can be factorized and resummed. \\
The particular mass cancellation mechanism occurring in the right-handed radiative decay is the consequence of the combination of two constraints: the angular momentum conservation in the pion vertex and the helicity flip in the photon-lepton vertex.\\
The situation is completely different if we consider the radiative process with the outgoing left-handed lepton, i.e. the process without helicity flip. The differential Inner Bremsstrahlung contribution for left-handed outgoing lepton is given by:
\begin{eqnarray}
& &\frac{1}{\Gamma_0}\frac{\diff \Gamma^L_{IB}}{\diff y} = 
\frac{\alpha}{4\pi} \, \frac{1}{(1-r)^2} \, \frac{1}{A(1-y+r)}
\bigg\{2A\, \big[r(2A-r+6) -y^2 -1-2A\big] \nonumber \\
& & + \big[(A-2r)(1+r)^2 + Ay(y-4r) -2ry(1+y) + y(5r^2+y^2+1)\big]
\, \ln{\frac{y+A}{y-A}} \nonumber \\
& & + (1-y+r)^2(2r-A-y) \, \ln{\frac{y+A-2}{y-A-2}} \bigg\}
\label{IBL}
\end{eqnarray}
The expression (\ref{IBL}) contains collinear singularities, since the coefficients of the collinear logarithms don't vanish in the limit \(r \to 0\), as one can see from eq. (\ref{IBL}). The decay rate (\ref{IBL}) is also infrared divergent, as one can verify by taking the limit \(y \to y^{MAX}\). In this case we don't have the constraint constituted by the helicity flip in the photon-lepton vertex and in the pion vertex the angular momentum is conserved for soft and virtual photon emission. Thus, in the left-handed case, the mass singularity cancellation
occurs in the ordinary way, as in the unpolarized case, i.e. in the total inclusive decay rate, obtained by adding all the order $\alpha$ contributions. \\
Let us show how the cancellation takes place. As we have already seen (eq. (\ref{bornpol})), in the Born \(\pi^{+}\) decay the outgoing lepton is left-handed, due to angular momentum conservation. 
Thus the unpolarized and the left-handed Born decay rate coincide:
\begin{equation}
\Gamma_0^L = \Gamma_0.
\label{0L0np}
\end{equation}
Because of the factorization with respect to the Born decay rate, also the unpolarized and the left-handed virtual contributions are equal:
\begin{equation}
\Gamma^L_v = \Gamma_v. 
\label{vLvnp}
\end{equation}
Expressing the left-handed Inner Bremsstrahlung contribution in terms of the unpolarized and the right-handed ones, the total contribution to order $\alpha$ to the left-handed process is given by:
\begin{equation}
\Gamma^L_{TOT} = (\Gamma_0 + \Gamma_v + \Gamma_{IB})
               - \Gamma^R_{IB}.
\label{TOTL}
\end{equation}
The expression (\ref{TOTL}) is finite both in the infrared and in the collinear limit, because the mass singularities present in the terms between brackets cancel each other, as we have seen (see eq. (\ref{tot}) and \(\Gamma^R_{IB}\) is free from mass singularities.\\
Let us now discuss the origin of these different cancellation mechanisms. The presence of mass singularities is a consequence of the fact that the states of a theory containing massless particles are highly degenerate. The KLN theorem states that the mass singularities disappear from the transition probability when we average it over the ensemble of degenerate states.
This theorem contains the Block-Nordsieck theorem as a special case, when we consider only the cancellation of the infrared divergences. \\
 We can define two degeneration ensembles, one relative to the infrared divergences and one relative to the collinear singularities. We call them the infrared and the collinear ensembles, respectively. If we sum the transition probability over the states contained in the former, the infrared divergences cancel out and if we average it over the latter, we obtain a quantity free from both infrared and collinear singularities. 
The infrared ensemble is the one prescribed by the  Block-Nordsieck theorem,
while the second is the one prescribed by the KLN theorem and contains the first as a subset. \\
Let us now examine how the infrared and collinear ensemble are composed for the radiative pion decay in the cases of unpolarized, left-handed and right-handed  outgoing lepton. This discussion concerns the issue of the degeneration of states already addressed for the unpolarized case \cite{Lee}. This issue in the case of helicity changing processes presents peculiar features. \\
We have seen in section \ref{pion} that in the unpolarized and left-handed Inner Bremsstrahlung contributions there are mass singularities, indicating that we have not summed the transition probability over the entire ensemble of degenerate states. In these cases the infrared ensemble contains, to order \(\alpha\), the state with a pion, a charged lepton and a neutrino and all the other states differing from this for the presence of a soft virtual or real photon, i.e. a photon with an energy \(E_{\gamma} < \omega\), where \(\omega\) is an infrared cut off, tipically the measurement apparatus resolution. 
The collinear ensemble is constituted by all the states of the infrared ensemble plus the states with a hard photon moving parallel to the pion or the lepton. Clearly the degeneration arises in the limit \(m_l \to 0\).\\
According to the KLN theorem the fact that \(\diff \Gamma^R_{IB}/\diff y\) is finite both in the infrared and in the collinear limits means that in the right-handed case, calculating the differential decay rate (i.e. summing over the photon polarization and integrating over the photon energy and emission angle), we have already averaged over the set of degenerate states relative to this process.
Let us now consider how the degeneration ensemble for the right-handed radiative decay is composed. To obtain \(\diff \Gamma_{IB}^R/\diff y\) we have not averaged over the infrared subspace of the collinear ensemble, but this is enough to render the transition probability free from mass singularities. Indeed in this case the infrared ensemble is empty, owing to the constraints imposed both by the angular momentum conservation in the pion vertex and by the helicity flip in the photon-lepton vertex. Thus in this case the degeneration ensemble contains only the states with the outgoing lepton accompanied by hard collinear photons.\\
We conclude that imposing to the outgoing lepton a polarization opposed to the one prescribed by the vertex preceding the photon emission, implies a reduction of the degeneration subspace. This fact has two consequences: the first is that both the infrared and the collinear limits are finite and the second is that these limits are disconnected, since the collinear degeneration subspace is constituted only by the states with the pion and the outgoing lepton accompanied by hard collinear photons. Thus in this case we have a particular application of the KLN theorem.  \\

\subsection{The $Z^0$ case}\label{Z}
In the radiative pion decay, due to the angular momentum conservation in the pion vertex, there is no room for a right-handed lepton. For such a channel, soft and virtual photon contributions are zero. This result is valid independently of the lepton mass.\\
Let us now consider a more general case, by loosing the value of the angular momentum of the decaying state. As an example, we study the radiative \(Z^0\) decay in a lepton-antilepton \((l^{-}l^{+})\) pair, in which the lepton is in a definite helicity state.\\
The \(Z^0\)-leptons vertex is:
\begin{displaymath}
i \, \frac{M_Z}{\sqrt{2}}\,\left(\frac{G}{\sqrt{2}}\right)^{1/2}\, 
\gamma_{\mu}(g_v - g_a \gamma_5),
\end{displaymath}
with
\begin{displaymath}
g_v = 1 - 4 \sin{\theta_W}^2 \;\;\;\;\;\;\;g_a=1
\end{displaymath}
where \(\theta_W\) is the Weinberg angle and \(M_Z\) is the \(Z^0\) mass.\\
We have chosen this process, since, by varying the constants \(g_v\) and \(g_a\), we can control the structure of the \(Z^0\)-leptons coupling; thus it is possible to point out the role played by the conservation law occurring in this vertex in the chiral limit in the collinear singularity cancellation. If we set \(g_v=g_a=1\), we require that in the limit of zero lepton mass, the \(Z^0\) couples to a left-handed lepton.\\
We calculate the decay rate for the process in which the lepton is right-handed. At the Born level this is given by:
\begin{eqnarray}
\Gamma^R_0 & = & \frac{GM^3_Z}{48\sqrt{2}\pi}\, \bigg\{\sqrt{1-4r}\,
\big[(g^2_v+g^2_a)(1-r) + 3r(g^2_v-g^2_a)\big]
- 2g_vg_a(1-4r) \bigg\}.
\label{bornz}
\end{eqnarray}
If in eq. (\ref{bornz}) we set \(g_v=g_a=1\), \(\Gamma^R_0\) vanishes in the chiral limit, since there isn't the term related to the axial anomaly.\\
Let us now study the decay process with the lepton emitting a real photon (see fig. 6)
\begin{figure}[t]
\begin{center}
\begin{tabular}{c}
\epsffile{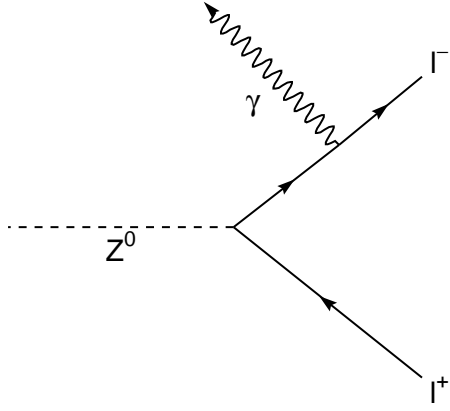}
\end{tabular}
\end{center}
\label{fig6}
\caption{Diagram corresponding to the decay rate \ref{Rz}.}
\end{figure}
and evaluate the probability that the outgoing lepton is right-handed. The electromagnetic interaction doesn't couples states with different chirality, hence the decay rate is expected to vanish for \(r \to 0\).\\
The decay rate for the process described by the diagram of fig. 6, differential with respect to the lepton energy, is given by:
\begin{eqnarray}
& &\frac{\diff \Gamma^R_{\gamma}}{\diff y} = 
\frac{\alpha G M^3_Z}{96 \sqrt{2}\pi^2}\,
\bigg\{\frac{(y-2)(1-y)^2}{4(1-y+r)^2}\,
\Big[(g^2_v+g^2_a)A + 2g_vg_a(2r-y)\Big] \nonumber \\
& + & \frac{(1-y)}{2(1-y+r)}\,
\Big[(g^2_v+g^2_a)A(2r-y) + 2g_vg_a(y^2-2r)\Big] \nonumber \\
& + & \frac{2}{(y-1)}\,
\Big[2(g^2_v-2g^2_a)Ar + (g^2_v+g^2_a)A + 2g_vg_a(4r-y^2+y-1)\Big] \nonumber \\
& + &\Big[g^2_v+g^2_a + 2g_vg_a\,\frac{1}{A}\,\big(4r^2+r(y-y^2+1)-y\big)\Big]
\,\Big[\ln{\frac{y+A}{y-A}} - \ln{\frac{y+A-2}{y-A-2}}\Big]\bigg\}. \nonumber \\
\label{Rz}
\end{eqnarray}
Here
\begin{displaymath}
r=\frac{m^2_l}{M^2_Z},
\end{displaymath}
\(y\) is the usual dimensionless variable:
\begin{displaymath}
y=\frac{2E_1}{M_Z}
\end{displaymath}
where \(E_1\) is the lepton energy and
\begin{displaymath}
A=\sqrt{y^2-4r}.
\end{displaymath}
The physical region for \(y\) is
\begin{equation}
2\sqrt{r} \le y \le 1.
\label{regionz}
\end{equation}
The result obtained, as given by the emission of the photon by a single leg, is gauge dependent. To have a gauge independent amplitude, the contribution of the diagram b) of fig. 7 must be added. For the purpose of the polarized amplitude, however, the helicity flip contribution of the diagram b) of fig. 7 gives zero in the massless limit and is therefore negligible in our discussion.\\
From now on we consider the case \(g_v=g_a=1\), to have the condition of chirality conservation in the \(Z^0\) vertex for \(m_l \to 0\). Taking this limit in eq. (\ref{Rz}), we see that \(\diff \Gamma^R_{\gamma}/\diff y\) does not vanish:
\begin{equation}
\frac{\diff \Gamma^R_{\gamma}}{\diff y} \longrightarrow 
\frac{\alpha}{2 \pi}\, (1-y) \Gamma_0 (Z^0 \to \nu \bar{\nu})
\;\;\;\;\;{\rm as}\;r \to 0,\;g_v=g_a=1.
\label{collz}
\end{equation}
The result of this limit is the contribution related to the axial anomaly, which has the same form of the one found in the pion case.\\
Let us now discuss the mass singularities cancellation mechanism for the \(Z^0\) decay case. If we keep the lepton mass different from zero, the helicity is not fixed by the interaction occurring before the photon emission, even if we set \(g_v=g_a=1\). Thus, for \(m_l \neq 0\), the soft and virtual photons contribution are different from zero and diverge in the infrared limit. Indeed, if we let the lepton energy reach its kinematical limit, \(y^{MAX}= 1\), we see that \(\diff \Gamma^R_{\gamma}/\diff y\) diverges. We expect the infrared divergences to cancel, if we add all the first order contributions, given by the diagrams of fig. 6 and 7 
\begin{figure}[t]
\begin{center}
\begin{tabular}{c}
\epsffile{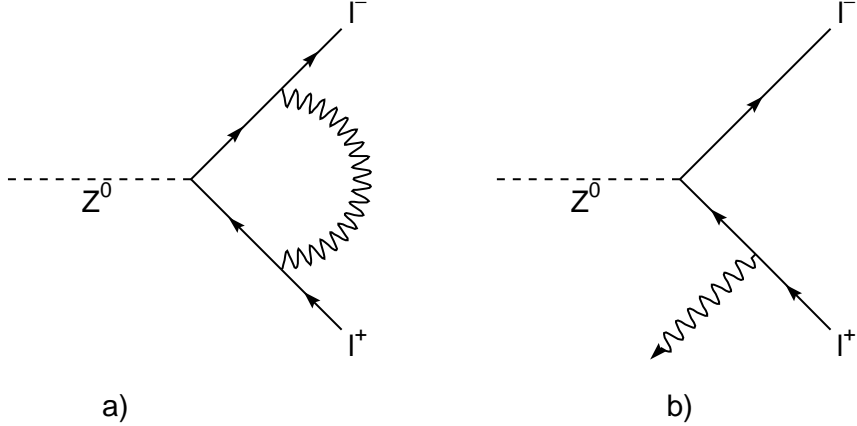}
\end{tabular}
\end{center}
\label{fig7}
\caption{The other order $\alpha$ diagrams}
\end{figure}
and calculate the totally inclusive decay rate.\\
Eq. (\ref{collz}) shows that the collinear limit gives a finite result. Thus, we conclude that, as in the case of the radiative pion decay, evaluating \(\diff \Gamma^R_{\gamma}/\diff y\) we have already summed over all the collinear degeneration subspace. As a consequence, the collinear and infrared limit are disconnected.\\
If we take the limit \(y \to 1\) in eq. (\ref{collz}), we obtain:
\begin{equation}
\frac{\diff \Gamma^R_{\gamma}}{\diff y} \longrightarrow 0
\;\;\;\;\;{\rm as}\;m_l\to 0, \;y \to 1 \;{\rm and}\;g_v=g_a=1.
\label{IRz}
\end{equation}
In general, if a quantity is finite in the collinear limit, it is finite also in the infrared limit, since the collinear subspace contains the infrared one. 
The eq. (\ref{IRz}) indicates that in the massless limit the soft photon contribution is zero. Indeed, it is factorized with respect to the Born decay rate, which, for \(r \to 0\) and \(g_v=g_a=1\), vanishes.  As we have discussed in section \ref{Anomaly}, the presence of the anomalous term is directly connected to the emission of the photon, hence it vanishes in the infrared limit.\\ 
The virtual photon contribution vanishes in the zero mass limit, as well. Indeed, it is factorized respect to the Born decay rate, which goes to zero as \(r \to 0\). The virtual correction factor can produce only a logarithmic collinear singularity, not a power-like one, needed for the cancellation of the chiral suppression.\\
We observe that taking the infrared limit \(x \to 0\) in eq. (\ref{hf}), gives a finite result (indeed the cross section vanishes). This is a consequence of the fact that the cross section (\ref{hf}) has been calculated to the leading order in the lepton mass. From the eq. (\ref{Rz}), we see that, for \(r \ll 1\), the infrared divergent term is given by:
\begin{equation}
\left ( \frac{\diff \Gamma^R_{\gamma}}{\diff y} \right)_{IR} \propto 
\frac{4r}{(y-1)} \left( \frac{2r}{y} - y - \frac{2}{y} \right) 
\label{IRd}
\end{equation}
and it is proportional to the lepton mass. Performing the calculation, neglecting the mass terms, as done in ref. \cite{Falk}, means imposing the chirality conservation law in the \(Z^0\) vertex; thus the soft photon contribution is zero and the infrared divergences disappear. To the leading order in the lepton mass, we have only the anomalous term, which vanishes in the infrared limit.\\
As done for the pion case, we now examine the composition of the degeneration ensembles for the radiative $Z^0$ decay with the right-handed lepton.\\
Eq. (\ref{IRz}) indicates that, in the chiral limit, the soft photons don't contribute to the process with the right-handed outgoing lepton, since, not carrying spin, they cannot contribute to the helicity 
flip.  \\
As we have already said, the virtual photon contribution vanishes in the zero mass limit.
In the limit \(m_l \to 0\), also the diagram with the photon emitted by \(l^{+}\) doesn't contribute, since, clearly, it violates the chirality conservation in the \(Z^0\) vertex. \\
The collinear degeneration arises in the massless limit. The result (\ref{collz}) shows that, in this limit, the collinear ensemble is constituted only by the states with the lepton accompanied by hard collinear photons, just as in the case of the pion decay. Thus for \(m_l \to 0\) the infrared ensemble is empty and also the states with the antilepton accompanied by a hard collinear photon don't contribute.\\

\section{Conclusions}\label{Conclus}
We have shown that the Dolgov and Zakharov treatment of the axial anomaly can be extended to processes characterized by a lepton which changes helicity by emitting a photon, as it was already noticed in \cite{S,Falk}. The corresponding decay rates don't vanish in the chiral limit, due to a term independent of the lepton mass; we interpret the presence of this term as related to the axial anomaly. This can be seen as a signal of the anomalous symmetry breaking in processes different from the usual ones, like the \(\pi^0 \to \gamma \gamma\) decay.\\
We have computed the rates corresponding to the \(\pi^{+}\) and \(Z^0\) radiative decays. We have analysed their infrared and collinear limits.  It results essential to keep the terms of all orders in the lepton mass, since the cancellation of the infrared and collinear divergences takes place among these terms. We have examined the connection between the polarization of the outgoing leptons and the application of the KLN theorem.\\
We have shown that for the helicity changing processes the cancellation of the collinear singularities occurs through a mechanism different from the usual one of real and virtual compensation. The coefficients in front of the collinear terms go to zero in the chiral limit, producing the finiteness of the distribution.  We have found, however, a difference between the pion case and the more general case of the \(Z^0\) decay. The former represents a particular case due to the angular momentum conservation in the pion vertex. As a consequence, the virtual and real soft photon contribution are zero, even if the lepton mass is kept different from zero. The Inner Bremsstrahlung contribution is finite both in the infrared and in the collinear limits.\\
In the \(Z^0\) case, the decay rate diverges in the infrared limit, since, for \(m_l \neq 0\), the soft photon contributions are not zero. However in the collinear limit, the result is finite, despite the fact that the collinear degenerate states arise only in the zero lepton mass limit. In this limit the virtual and real soft photon contributions do vanish. In order to make the collinear limit finite, it is therefore sufficient to sum over degenerate states made of the changing helicity lepton accompanied by a hard collinear photon. The transition probability becomes finite after summing over the photon final phase space.\\
We noticed that in the helicity changing processes the collinear limit results disconnected from the infrared one. The contributions coming from the virtual photon emission and from the emission of photons by particles different from the one changing helicity, are zero.\\
This situation is due to the fact that the Born part of the process fixes the fermion chirality in the zero mass limit, while, after the photon emission, it is in a state of opposite chirality; this reduces the collinear ensemble.\\
We conclude that the collinear singularity cancellation mechanism for helicity changing processes is controlled by the anomalous breaking of the chiral symmetry. The axial anomaly implies that the collinear limit gives a finite result, independent of the fermion mass.\\
The extension of the above remarks to other gauge theories like QCD, is possible. It could allow a more systematic and complete treatment of the infrared and collinear singularities. \\




\vspace{3mm}
\begin{center}
\begin{large}
{\bf Acknowledgments}
\end{large}
\end{center}
\noindent We wish to thank V. Fadin and E. Kuraev for valuable comments and S. Forte, J. Kodaira and L. Lipatov for useful discussions.

\end{document}